\documentclass
[preprint,byrevtex,showpacs,balancelastpage,prl,10pt,twocolumn,a4paper]{revtex4}%
\usepackage{amsfonts}
\usepackage{amsmath}
\usepackage{amssymb}
\usepackage[dvips]{graphicx}%
\begin{document}
\title{Intrinsic electron-glassiness in strongly-localized Be films}
\author{Z. Ovadyahu$^{1}$, X. M. Xiong$^{2}$ and P. W. Adams$^{2}$ }
\affiliation{$^{1}$Racah Institute of Physics, The Hebrew University, Jerusalem, 91904,
Israel, $^{2}$Department of Physics and Astronomy, Louisiana State University,
Baton Rouge, Louisiana 70803.}

\begin{abstract}
We present results of out--of-equilibrium transport measurements made on
strongly-localized Beryllium films and demonstrate that these films exhibit
all the earmarks of intrinsic electron-glasses. These include slow
(logarithmic) relaxation, memory effects, and more importantly, the
observation of a memory dip that has a characteristic width compatible with
the carrier-concentration of beryllium. The latter is an empirical signature
of the electron-glass. Comparing various non-equilibrium attributes of the
beryllium films with other systems that exhibit intrinsic electron-glasses
behavior reveals that high carrier-concentration is their only common feature
rather than the specifics of the disorder that rendered them insulating. It is
suggested that this should be taken as an important hint for any theory that
attempts to account for the surprisingly slow relaxation times observed in
these systems.

\end{abstract}
\pacs{\bigskip72.20.Ee 72.20.Ht 72.70.+m}
\maketitle

\section{Introduction}

The interplay between static disorder and Coulomb interactions may precipitate
a glassy state in an Anderson insulator. This `electron-glass' scenario was
discussed in several papers \cite{1,2,3,4}. In theory, this property is
generic to \textit{all} degenerate Fermi systems with localized states
interacting via a Coulomb potential. Experimental evidence for these glassy
effects, however, has been somewhat scarce, presumably due to specific
material requirements. It turns out that only systems with relatively high
carrier-concentration $n$ exhibit relaxation times that can be conveniently
monitored by transport measurements. Conductance relaxations that persist for
many seconds, and memory effects characteristic of \textit{intrinsic }\cite{5}
electron-glass, seem to be peculiar to systems with $n>10^{19}$cm$^{-3}$
\cite{5}. A prominent group of materials that exhibit electron-glass behavior
with long relaxation times are granular metals; Al \cite{6}, Bi \cite{7}, Pb
\cite{7}, Ni \cite{8}, and Au \cite{9}, all having high carrier-concentration
$n\eqslantgtr10^{22}$cm$^{-3}.$

Hitherto, the only non-granular systems that exhibited intrinsic
electron-glass behavior were crystalline and amorphous indium-oxide films
(In$_{2}$O$_{3-x}$ and In$_{x}$O respectively) \cite{5}, which are ionic compounds.

In this work we report on the low temperature transport properties of
strongly-localized Be films, and demonstrate that they exhibit intrinsic
glassy effects. This is the first non-granular mono-atomic system to show
these effects. These include logarithmic relaxation of the out-of-equilibrium
conductance and, more importantly, a memory-dip that has all the earmarks of
intrinsic electron-glass. Although beryllium, like a typical metal, has a
Fermi energy E$_{F}$ of few electron-volts, it has an unusually low density of
states at E$_{F}$ \cite{10}. A signature of the Be peculiar density of states,
namely, at the Fermi energy, the density-of-states decreases with energy, is
actually observed in our field effect measurements as will be demonstrated
below. The low density of states of beryllium is presumably the main reason
why strong localization is achievable in this material just by making the
sample thin enough (yet still physically continuous).

\section{Experimental}

Samples used in the experiments reported here were 18$\pm$5\AA ~films
deposited as described elsewhere \cite{11} on 140$\mu$m glass slides. These
were silver-painted on their backside so as to form a gate for the field
effect measurements. The samples were typically 400$\mu$m wide and 600$\mu$m
long strips and had sheet resistance R$_{\square}$ in the range 23k$\Omega$ to
120k$\Omega$ at 295K and 100k$\Omega$-160M$\Omega$ at $\approx$4K. The
strongly-localized nature of the films at this temperature range was tested by
measuring their conductance versus temperature dependence G(T) in the range
4-50K. Below$\approx$10K, this dependence is consistent with G(T)$\propto\exp
$[-($\frac{T_{0}}{T}$)$^{1/2}$] with T$_{0}$ in the range 100-900K. Figure 1
illustrates this behavior for two of the samples that were used in this study.%

\begin{figure}
[ptb]
\begin{center}
\includegraphics[
trim=0.000000in 0.513428in 0.000000in -0.286134in,
height=3.32in,
width=3.3157in
]%
{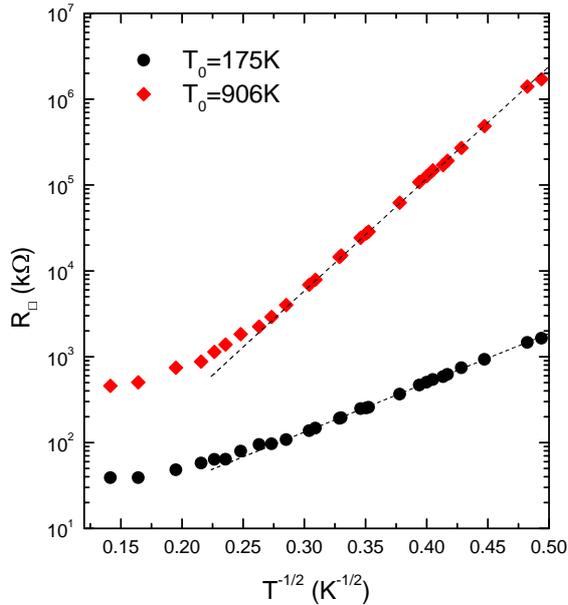}%
\caption{Resistance versus temperature plots for two of the beryllium films
used in this study. }%
\end{center}
\end{figure}
The different values of R$_{\square}$ in the studied series of samples were
obtained by a judicious oxidation of the Be films in an oxygen-enriched
chamber. The change of the samples resistance was constantly monitored during
the oxidation process. At the range of sample thickness d $\approx$18
\AA ,~the room-temperature sheet-resistance of the samples is $\approx
$20k$\Omega$. Such samples are usually deep into the hopping regime at liquid
helium temperatures, and a small change in their thickness d translates into a
large change of R$_{\square}$ at 4K. The sheet-resistance of samples much
thicker than 18~\AA ~may not reach the quantum resistance at 4K (recall that
R$_{\square}$%
$>$%
$\hbar$/e$^{2}$ is a pre-requisite for electron-glass behavior \cite{5}). For
example, in a previous study \cite{12} a beryllium film with a nominal
thickness d of $\approx$20$~$\AA ,~only slightly thicker than the films
studied here, exhibited G(T)$\propto\exp$[-($\frac{T_{0}}{T}$)$^{1/2}$] with
T$_{0}$=1.6K. This should be compared with T$_{0}\gtrsim$100K in the samples
used in this work. The much smaller value for T$_{0}$, and the associated
larger localization length in this case means that strongly-localized behavior
is attained only at temperatures that are well below the range covered here.
Such samples are not included in our present study where one of our goals is
to compare the results with previously studied electron-glasses, which were
measured at or near 4K.

The conductivity of the samples was measured using a two-terminal ac technique
employing a 1211-ITHACO current pre-amplifier and a PAR-124A lock-in
amplifier. Measurements reported below were performed with the samples
immersed in liquid helium at T=4.1K maintained by a 100 liters storage-dewar,
which allowed long term measurements of samples as well as a convenient way to
maintain a stable temperature bath. Unless otherwise indicated, the ac voltage
bias was small enough to ensure linear response conditions (judged by Ohm's
law being obeyed within the experimental error). Fuller details of
measurements techniques are given elsewhere \cite{13}.

A variety of techniques were employed to characterize the films
microstructure. Fig.2 shows a Transmission Electron Microscope (TEM)
micrograph of a Be film prepared in the same way and with similar thickness as
the samples used for the transport studies.%
\begin{figure}
[ptb]
\begin{center}
\includegraphics[
trim=0.000000in 0.000000in 0.010366in 0.000000in,
height=2.3514in,
width=3.3122in
]%
{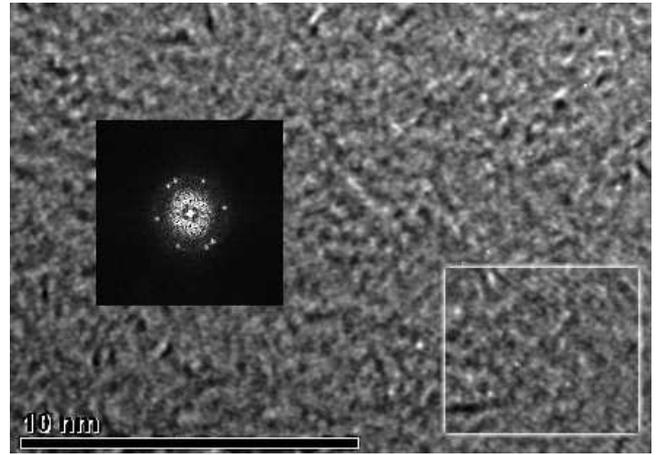}%
\caption{Transmission Electron Microscope micrograph of $\approx$20\AA ~Be
film deposited on carbon-coated Cu grid. The black square is a Fourier
transform of the framed area on the right. The latter shows 2.33\AA ~fringes
associated with the d-spacing of the $<$100$>$ planes of the hexagonal BeO
(reflected as satellites in the Fourier transform plate).}%
\end{center}
\end{figure}
The physical continuity of the film is evident in the figure. On careful
examination, the micrograph shows occasional fringes that indicate the
presence of small crystals. These were identified as BeO by direct imaging and
further, by their diffraction pattern (interestingly, the diffraction pattern
of Be, being light on electrons, was presumably too weak to register a clear
pattern over the background set by the amorphous carbon support-film). The BeO
crystals were clearly observable in TEM dark-field imaging off their
$<$%
100%
$>$
diffraction line. This enabled an estimate of their crystallographic size and
partial volume in the samples. Randomly distributed BeO crystallites of sizes
up to 50-70\AA ~were observed in these dark-field images. We estimate that
less than $\approx$10\% of the film area is occupied by fully oxidized Be
crystallites, and therefore transport presumably occur through non-oxidized Be
phase. Yet, the insulating BeO crystallites, somewhat restrict the volume
available for conductivity (much like punching holes in the film would). A
result of this geometrically-constrained structure is that the transport
properties of the films show some mesoscopic effects that one usually
encounters in smaller systems measured at similar temperatures and comparable
degree of disorder \cite{14,19}.

The physical continuity of the Be phase in the film was ascertained by
performing local electron-energy-loss-spectroscopy (EELS) on the parts of the
film that were not occupied by BeO crystals. The EELS spectra taken from
theses areas was consistent with that of metallic Be. A slight shift of
energy, $\eqslantless$+3\% of the peaks position in the spectra was detected,
possibly due to strain. The presence of free Be in the samples was also
confirmed by X-ray Photoemission Spectroscopy, which was carried out on the
actual samples that were used for the transport measurements.

\section{Results and discussion}

We turn now to the non-equilibrium transport properties of the films. The
first signature of glassy behavior in these films is encountered upon
quench-cooling the sample to 4K; after an initial fast drop (reflecting the
change in temperature), the conductance G keeps on decreasing slowly
(logarithmically) long after the sample has reached the bath temperature. A
typical quench-cooling protocol is shown in Fig.3a.
\begin{figure}
[ptb]
\begin{center}
\includegraphics[
trim=0.000000in 0.451973in 0.618209in -0.204092in,
height=3.371in,
width=3.3122in
]%
{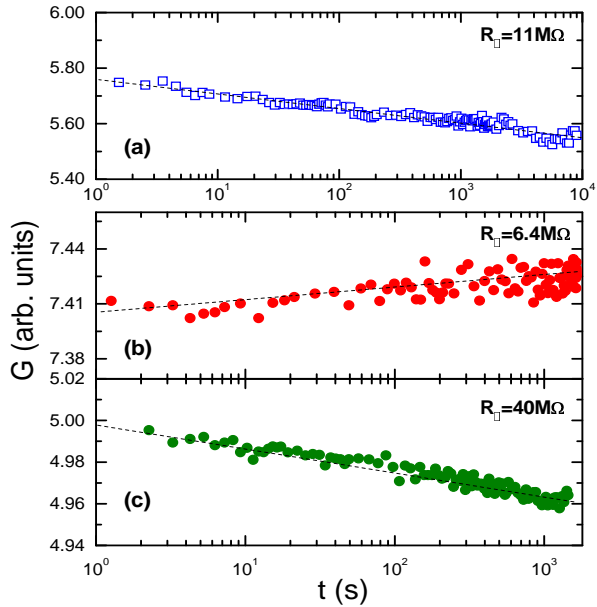}%
\caption{Non-equilibrium transport behavior of typical Be films under
different protocols; (a) After a quench-cool from T$\approx$100K to T=4.11K.
(b) Starting from near-equilibrium, G(t) was recorded from the instance the
source-drain voltage was increased to 10V$_{ac}$ from the (Ohmic) 20mV$_{ac}$
bias. (c) Same sample as (b) after the 20mV$_{ac}$ has been restored. Each
graph is labeled with the average R$_{\square}$of a sample under the
measurements conditions.}%
\end{center}
\end{figure}
The figure also illustrates the slow conductance excitation process upon
`stressing' \cite{15} the film with a non-Ohmic source-drain voltage Fig.3b.
The ensuing relaxation of G after the source-drain voltage was set back to its
Ohmic value is shown in Fig.3c. These excitation-relaxation curves are clearly
similar to those previously observed in glassy In$_{2}$O$_{3-x}$ samples
\cite{15}

A controlled way to take the system out of equilibrium is a change of the
potential difference between the sample and a near-by gate. This technique has
been widely used in the study of several electron-glasses \cite{5,6,7,13}.
Among other things, it may be used to estimate a typical relaxation time
$\tau~$under a given set of conditions \cite{16}. An example of such protocol
is illustrated in Fig.4.%
\begin{figure}
[ptb]
\begin{center}
\includegraphics[
trim=0.000000in 0.677959in 0.228101in 0.459011in,
height=2.9758in,
width=3.4835in
]%
{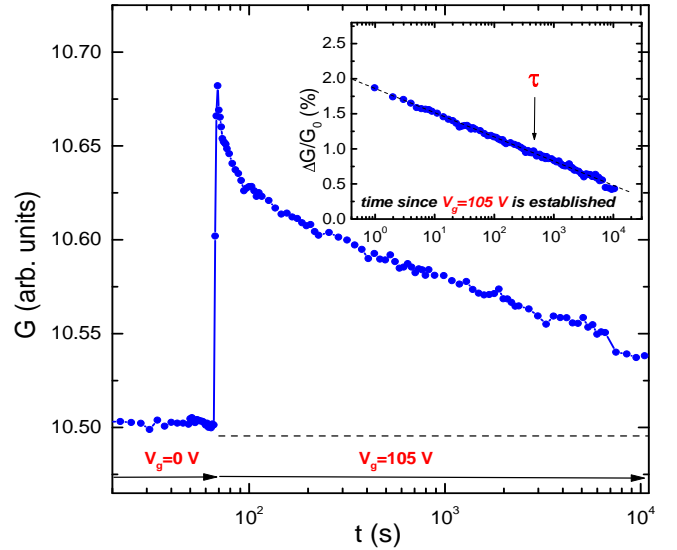}%
\caption{A gate-excitation protocol for a specific Be sample (R$_{\square}%
=$40M$\Omega).$ The insert depicts the characteristic log(t) dependence of the
EG from which the typical relaxation time $\tau$ is estimated using the
baseline conductance G$_{0}$(105V) (marked here by the dashed line).measured
independently.}%
\end{center}
\end{figure}
In this protocol, one uses the conductance relaxation law $\Delta$G(t/t$_{0}%
$)$=\Delta$G$(1\sec)-a\cdot\log$(t/t$_{0}$) where t$_{0}$ is the experimental
resolution time, and the equilibrium value of G at V$_{g}$=105V to extract the
value of $\tau$ defined by G$(\tau)\equiv\frac{1}{2}$G$(1\sec).$

A characteristic feature that is believed to be common to all intrinsic
electron-glasses is a memory-dip; this is a cusp-like minimum in G(V$_{g}$)
centered at the gate voltage V$_{g}$ where the system was allowed to
equilibrate \cite{5,6,7,17}. A conspicuous memory-dip was consistently
observed in all our Be films. Fig.5 shows this feature for two Be samples in
the studied series. For both samples, G(V$_{g}$) scans were taken after a
$\simeq$24 hour equilibration under V$_{g}$=0 volt. Note first that the
memory-dips have the same shape and width independent of R$_{\square}$ and
independent of wether the G(V$_{g}$) scans were taken by sweeping V$_{g}$
\textit{through }the equilibrium-V$_{g}$ or \textit{symmetrically} around it
\cite{13} (c.f., the lower graph of Fig.5).%
\begin{figure}
[ptb]
\begin{center}
\includegraphics[
trim=0.000000in 0.535997in 0.148572in 0.000000in,
height=3.2465in,
width=3.3122in
]%
{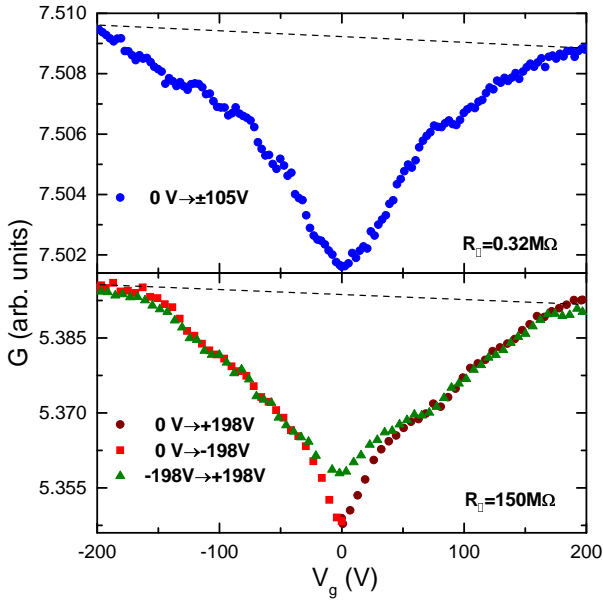}%
\caption{The conductance versus gate voltage for two Be films (labeled by
their R$_{\square})$ illustrating the memory-dip structure. After the sample
was allowed to equilibrate under V$_{g}$=0 for $\approx$24hours, scans were
taken either from V$_{g}$=0 in each voltage direction (symmetrical G(V$_{g}$)
scans) or, from -198V to +198V. Note that the structure is skewed due to the
contribution of the anti-symmetric (equilibrium) field-effect (see text).}%
\end{center}
\end{figure}

In previously studied electron-glasses the width of the memory-dip was found
to systematically depend on the carrier-concentration $n$ of the system
\cite{17}. On the basis of the G(V$_{g}$) data, we have estimated the typical
width in our Be films in the same manner as was done in \cite{17}. This
involves several stages; First the change of charge $\Delta$Q associated with
the cusp-width is estimated from $\Delta$V$_{g}$ by taking heed of the
sample-gate capacitance. The relevant energy is then calculated using the
beryllium ($\partial$n/$\partial\mu$)$_{E_{F}}$ and the screening length. This
procedure gave the energy-width $\Gamma^{\ast}$ (as defined in \cite{17}) as
$\approx$8meV. Using the empirical relation between the width and $n$ (Fig.4
of \cite{17}) such $\Gamma^{\ast}$ corresponds to $n$ of order 10$^{21}%
$-10$^{22}$cm$^{-3}$. This is consistent with our Hall effect measurements
that gave $n$ $\cong$7-8$\cdot$10$^{21}$cm$^{-3}$ as well as with the
concentration predicted by band-structure calculations ($n$ = 0.016 state/atom
\cite{18} tantamount to $n\simeq1.7\cdot10^{21}$cm$^{-3}$). This correlation
between the width of the memory-dip and the carrier concentration of the
material is an important empirical test for the intrinsic nature of the
electron-glass \cite{5}.

The G(V$_{g}$) traces (Fig.5) reveal some mesoscopic fluctuations
(reproducible with V$_{g}$ scans) superimposed on the memory-dip (note e.g.,
the modulation of G(V$_{g}$) around V$_{g}\approx-$120V, and +90 for the
0.32M$\Omega$ and the 150M$\Omega$ samples respectively). As alluded to above,
these presumably result from the small effective volume for conductance due to
the presence of the insulating BeO regions.

The relative magnitude of the memory-dip $\Delta$G/G grows monotonously with
R$_{\square}$. Interestingly, $\Delta$G/G(R$_{\square}$) for Be is almost
identical to that measured under the same conditions in other
electron-glasses. In Fig.6 we compare the results of the current study with
some old data \cite{19} taken on In$_{2}$O$_{3-x}$ films exhibiting quite
similar behavior.%
\begin{figure}
[ptb]
\begin{center}
\includegraphics[
trim=0.000000in 0.000000in 0.231325in 0.000000in,
height=3.4402in,
width=3.4411in
]%
{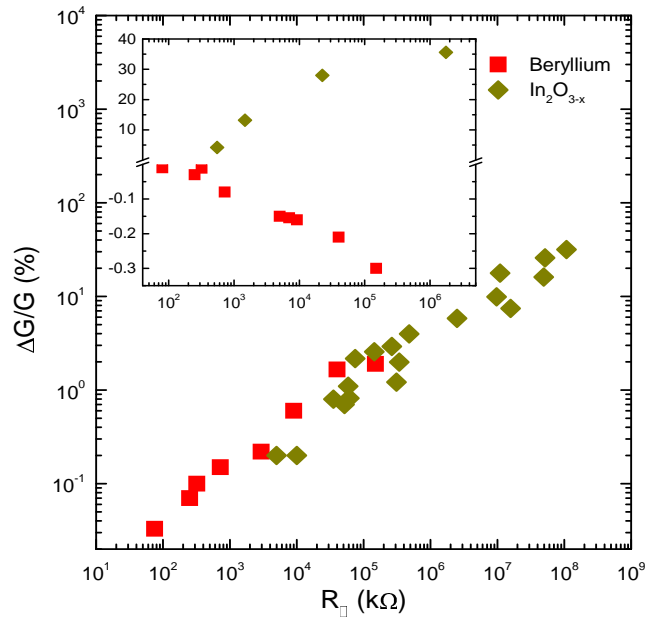}%
\caption{The relative magnitude of the memory-dip versus R$_{\square}$ for the
studied Be samples. The data of \cite{15} are shown for comparison. Insert:
The slope (relative change of G per 400V of V$_{g}$) of the anti-symmetric
part of G(V$_{g}$) compared with typical results for In$_{2}$O$_{3-x}%
$~samples.}%
\end{center}
\end{figure}
A similar agreement is observed between that data of Fig.6 and the results
obtained on granular aluminum films (c.f., Fig.17 of \cite{20}).

On the other hand, the anti-symmetric part (c.f., Fig.6) of the G(V$_{g}$) for
the beryllium samples has the \textit{opposite} slope to that observed in
In$_{2}$O$_{3-x}$ and In$_{x}$O \cite{5,13}. The sign of this slope is
controlled by the energy derivative of the thermodynamic density of states
$\partial$n/$\partial\mu$ at the Fermi level; $\partial$G(V$_{g}$)/$\partial
$V$_{g}\propto\frac{\partial}{\partial E}$($\partial$n/$\partial\mu$)$_{E_{F}%
}$. The anti-symmetric part of $\partial$G(V$_{g}$)/$\partial$V$_{g}$ we
observe in all our Be films (represented by the dashed curves in Fig.5) is
consistent with the negative slope of $\partial$n/$\partial\mu$%
$\vert$%
$_{E_{F}}$ found in theoretical calculations for the Be band-structure
\cite{21}. These calculations assumed an ideal Be crystal which might not be
relevant for the disordered structure. It is not uncommon however that
band-structure features calculated for the perfect crystal persist in the
disordered material (as actually observed for some optical properties of Be
\cite{22}). The magnitude of the slope depends also on the film resistance as
is shown for both In$_{2}$O$_{3-x}$ and Be in the insert to Fig.6. Being a low
density system, $\frac{\partial}{\partial E}$($\partial$n/$\partial\mu
$)$_{E_{F}}$ in In$_{2}$O$_{3-x}$~is much larger than in metals, which in turn
makes $\partial$G(V$_{g}$)/$\partial$V$_{g}$ larger.

It is interesting to note that the currently known electronic systems that
exhibit intrinsic glassiness (with associated long relaxation times) are quite
diverse in most other aspects. For example, in terms of microstructure, there
are in this group representatives of all types of disordered structures;
poly-crystalline (In$_{2}$O$_{3-x}$, and Be), granular or discontinuous (Al,
Pb, Au, Ni), and amorphous (In$_{x}$O, Bi). Most of these systems contain
oxide, whether as an intrinsic part of the material (In$_{2}$O$_{3-x},$%
In$_{x}$O), or to stabilize a granular structure (e.g., Al). However, the
discontinuous Ni films, being prepared and measured under high-vacuum
conditions \cite{8}, is oxygen-free and show the same effects as the other
intrinsic electron-glasses \cite{8}. In some of these systems there may be
local order due to superconductivity (Pb, Bi, the high n version of In$_{x}$O,
Be, Al) or magnetism (Ni) at the temperatures of the experiments but not in
others. Finally, all these systems obey some form of activated conductivity,
G(T)$\propto\exp$[-($\frac{T_{0}}{T}$)$^{\alpha}$] with 0.3%
$<$%
$\alpha$%
$<$%
1, however, no single value of $\alpha$ is singled out in the group. In other
words, the conductivity versus temperature law G(T) exhibited by these systems
is not due to a specific hopping mechanism. It does not set this group apart
from other hopping systems that do not exhibit long relaxation times. Indeed,
a G(T) law that resembles is observed in many disordered semiconductors while
their relaxation times are very short \cite{23} (presumably due to their low
$n$ \cite{5}).

In fact, the \textit{only} common feature of the materials that show long
relaxation times appears to be their relatively high carrier-concentration (in
addition of course to being strongly-localized thus exhibiting hopping
conductivity). The common, out-of-equilibrium features that all these
electron-glasses exhibit are suggestive of a generic mechanism. The current
work supports the conjecture that sluggish relaxation and the associated
memory effects of the electron-glass is intimately connected with high
carrier-concentration \cite{5}. This should be the pivotal point of any
theoretical model that purports to account for these phenomena.

ZO and PWA gratefully acknowledge discussions with the participants of the
Electron-Glass program (organized by the Kavli Institute of Theoretical
Physics) during which this manuscript was prepared. This research was
supported by a grant administered by the US Israel Binational Science
Foundation and by the Israeli Foundation for Sciences and Humanities. PWA
acknowledges the support of the US Department of Energy under Grant No.\ DE-FG02-07ER46420.


\begin{thebibliography}{99}                                                                                               %


\bibitem {1}J. H. Davies, P. A. Lee, and T. M. Rice, Phys. Rev. Lett.,
\textbf{49}, 758 (1982).

\bibitem {2}M. Gr\"{u}newald, B. Pohlman, L. Schweitzer, and D. W\"{u}rtz, J.
Phys. C, \textbf{15}, L1153 (1982).

\bibitem {3}M. Pollak and M. Ortu\~{n}o, Sol. Energy Mater., \textbf{8}, 81
(1982); M. Pollak, Phil. Mag. B\textbf{ 50}, 265 (1984).

\bibitem {4}Ariel Amir, Yuval Oreg, and Yoseph Imry, Phys. Rev. B \textbf{77},
165207 (2008).; Phys. Rev. Lett. \textbf{103}, 126403 (2009).

\bibitem {5}Z. Ovadyahu, Phys. Rev. B \textbf{78}, 195120 (2008).

\bibitem {6}T. Grenet, Eur. Phys. J, \textbf{32}, 275 (2003); T. Grenet, J.
Delahaye, M. Sabra, and F. Gay, Eur. Phys. J. B \textbf{56}, 183 (2007).

\bibitem {7}G. Martinez-Arizala, D. E. Grupp, C. Christiansen, A. Mack, N.
Markovic, Y. Seguchi, and A. M. Goldman, Phys. Rev. Lett., \textbf{78}, 1130
(1997). G. Martinez-Arizala, C. Christiansen, D. E. Grupp, N. Markovic, A.
Mack, and A. M. Goldman, Phys. Rev. B\textbf{ 57}, R670 (1998).

\bibitem {8}A. Frydman (in private communication).

\bibitem {9}C. J. Adkins, J. D. Benjamin, J. M. D. Thomas, J W Gardner, and A.
J. McGeown J. Phys. C: Solid State Phys. \textbf{17,} 4633 (1984). Note
however that no time dependence of the effects has been reported in this paper.

\bibitem {10}Guenter Ahlers, Phys. Rev. \textbf{145}, 419 (1966).

\bibitem {11}P. W. Adams, P. Herron, and E. I. Meletis, Phys. Rev. B,
\textbf{58}, R2952 (1998).

\bibitem {12}V. Yu. Butko, J. F. DiTusa, and P. W. Adams, Phys. Rev. Lett.,
\textbf{84}, 1543 (2000); V.Yu. Butko and P.W. Adams, Nature \textbf{409}, 161 (2001).

\bibitem {13}A. Vaknin, Z. Ovadyahu, and M. Pollak, Phys. Rev. B\textbf{ 65},
134208 (2002).

\bibitem {14}V. Orlyanchik, and Z. Ovadyahu, Phys. Rev. B \textbf{75}, 174205 (2007).

\bibitem {15}V. Orlyanchik, A Vaknin, and Z. Ovadyahu, Phys. Stat. Sol.,
B\textbf{ 230}, 67 (2002); V. Orlyanchik, and Z. Ovadyahu, Phys. Rev. Lett.,
\textbf{92}, 066801 (2004).

\bibitem {16}Z. Ovadyahu, Phys. Rev. B \textbf{73}, 214208 (2006).

\bibitem {17}A. Vaknin, Z. Ovadyhau, and M. Pollak, Phys. Rev. Lett.,
\textbf{81}, 669 (1998).

\bibitem {18}J. H. Tripp, P. M. Everett, W. L. Gordon, and R. W. Stark, Phys.
Rev., \textbf{180}, 669 (1969).

\bibitem {19}M. Pollak and Z. Ovadyahu, J. de Physique I France, \textbf{7},
1595 (1997).

\bibitem {20}J. Delahaye, T. Grenet and F. Gay, Eur. Phys. J. B \textbf{65}, 5 (2008).

\bibitem {21}E. Wimmer, J. Phys. F: Met. Phys. \textbf{14,} 681 (1984); G.
Pari, Vijay Kumar, A. Mookerjee, and A. K. Bhattacharyya, J. Phys.: Condens.
Matter \textbf{11, }4291 (1999); R. Bo\~{c}a, P. Hajko, and L. Benko,
Czechoslovak J. of Physics, \textbf{44}, 897 (1994).

\bibitem {22}O. Hunderi and H. .P Myers, J. Phys. F: Metal Phys., \textbf{4},
1088 (1974).

\bibitem {23}Don Monroe A. C. Gossard, J. H. English, B. Golding, W. H.
Haemmerle, and M. A. Kastner, Phys. Rev. Lett. \textbf{59}, 1148 (1987); V.K.
Thorsm\o lle, and N.P. Armitage, Phys. Rev. Lett. 105, 086601 (2010).
\end{thebibliography}
\end{document}